%% file: Main.tex
\documentclass[aps,twocolumn,superscriptaddress,amsmath,longbibliography]{revtex4-2}
\usepackage[pdftex]{graphicx,color,hyperref}
\usepackage{soul}
\usepackage{amsfonts}
\usepackage{amssymb}
\usepackage{amsmath}
\usepackage{amsthm}
\usepackage{bm}
\usepackage{xcolor}
\usepackage{braket}
\usepackage{enumerate}
\usepackage[capitalise]{cleveref}
\usepackage[normalem]{ulem}
\usepackage{amsmath,epsfig,color,amssymb}
\usepackage{graphicx}
\usepackage{xcolor}
\input{commands2018}

\begin{document}

\title{Singularities in nearly-uniform 1D condensates due to quantum diffusion}

\author{C. L. Baldwin}
\affiliation{Joint Quantum Institute, NIST/University of Maryland, College Park, Maryland 20742, USA}
\author{P. Bienias}
\affiliation{Joint Quantum Institute, NIST/University of Maryland, College Park, Maryland 20742, USA}
\affiliation{Joint Center for Quantum Information and Computer Science, NIST/University of Maryland, College Park, Maryland 20742, USA.}
\author{A. V. Gorshkov}
\affiliation{Joint Quantum Institute, NIST/University of Maryland, College Park, Maryland 20742, USA}
\affiliation{Joint Center for Quantum Information and Computer Science, NIST/University of Maryland, College Park, Maryland 20742, USA.}
\author{M. J. Gullans}
\affiliation{Joint Center for Quantum Information and Computer Science, NIST/University of Maryland, College Park, Maryland 20742, USA.}
\author{M. Maghrebi}
\affiliation{Department of Physics and Astronomy, Michigan State University, East Lansing, Michigan 48824, USA}

\date{\today}

\begin{abstract}

Dissipative systems often exhibit wavelength-dependent loss rates.
One prominent example is Rydberg polaritons formed by electromagnetically-induced transparency, which have long been a leading candidate for studying the physics of interacting photons and also hold promise as a platform for quantum information.
In this system, dissipation is in the form of quantum diffusion, i.e., proportional to $k^2$ ($k$ being the wavevector) and vanishing at long wavelengths as $k\to 0$.
Here, we show that one-dimensional condensates subject to this type of loss are unstable to long-wavelength density fluctuations in an unusual manner: after a prolonged period in which the condensate appears to relax to a uniform state, local depleted regions quickly form and spread ballistically throughout the system.
We connect this behavior to the leading-order equation for the nearly-uniform condensate---a dispersive analogue to the Kardar-Parisi-Zhang (KPZ) equation---which develops singularities in finite time.
Furthermore, we show that the wavefronts of the depleted regions are described by purely dissipative solitons within a pair of hydrodynamic equations, with no counterpart in lossless condensates.
We close by discussing conditions under which such singularities and the resulting solitons can be physically realized.
    
\end{abstract}

\maketitle

Dissipative systems are typically described by a constant dissipation rate, yet many physical platforms are instead subject to momentum-dependent losses.
A prominent example is Rydberg systems, which have received much interest as a platform for quantum nonlinear optics~\cite{Gorshkov2011Photon,Peyronel2012Quantum,Dudin2012Strongly} and quantum information processing/simulation~\cite{Friedler2005LongRange,Saffman2010Quantum,Weimer2010Rydberg,Li2013Entanglement,Maxwell2013Storage,Otterbach2013Wigner,Gorniaczyk2016Enhancement,Tiarks2016Optical,Bienias2020Two}.
The polaritons that form under the condition of electromagnetically-induced transparency (EIT)~\cite{Harris1990Nonlinear,Fleischhauer2005Electromagnetically,Mohapatra2007Coherent} undergo quantum diffusion, i.e., a one-body loss rate $\Gamma_k \propto k^2$~\cite{Peyronel2012Quantum}.
A similar form of dissipation occurs in bosonic atoms driven by two coherent laser beams~\cite{Fleischhauer2005Electromagnetically}.
This type of loss can be realized in arrays of microwave resonators as well by coupling the cavity modes to qubits~\cite{Marcos_2012,Marino2016Driven}.

In a many-body system, momentum-dependent loss can have drastic consequences, from dissipatively stabilizing condensates~\cite{Diehl2008Quantum} to producing exotic critical or correlated states~\cite{eisert2010noise, Honing_2012, Marino2016Driven, Zeuthen_2017}.
These advances notwithstanding, many consequences of momentum-dependent loss remain undiscovered.

\begin{figure}[h!]
\centering
\includegraphics[width=1.0\columnwidth]{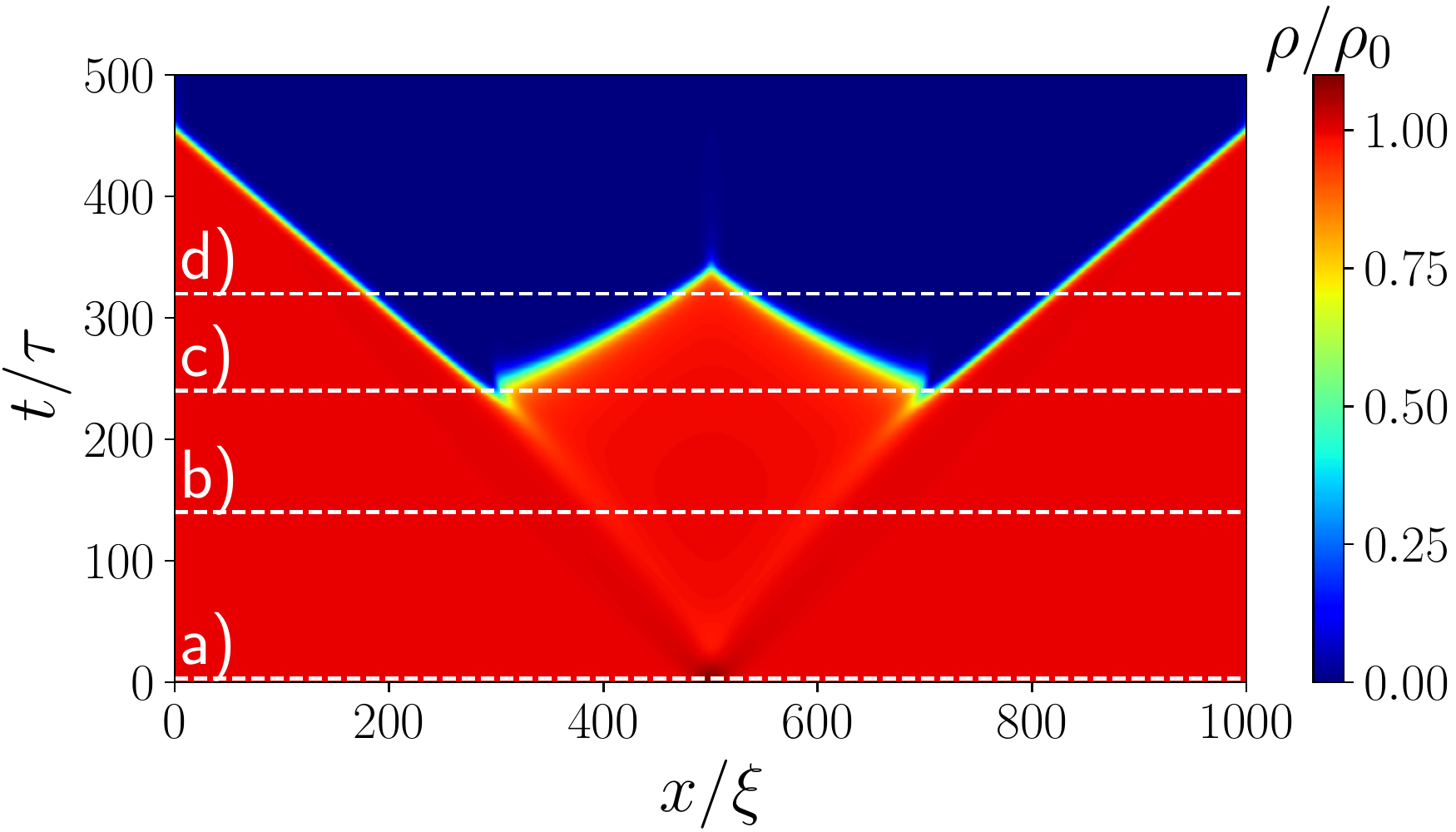}
\includegraphics[width=1.0\columnwidth]{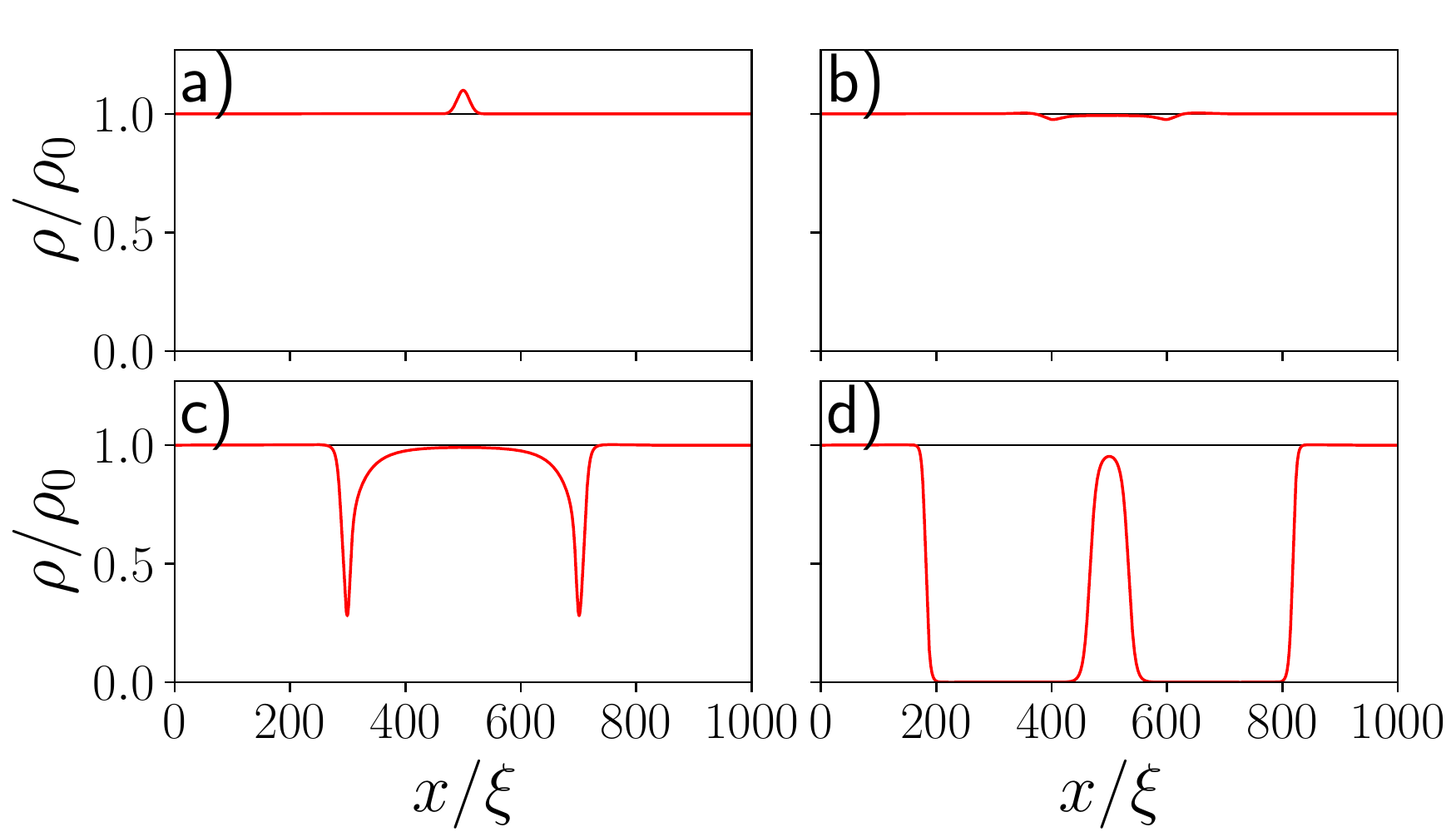}
\caption{(Top) Time evolution of a small Gaussian density perturbation. (Bottom) Snapshots of the density profile (normalized by the initial density $\rho_0$) at the times indicated by the dashed white lines in the top panel. Simulation parameters are the dissipation strength $\lambda = 2.0$, height of the initial Gaussian $h = 0.1$, its width $w = 15 \xi$, the spatial discretization $\Delta x = 0.2 \xi$, and the time step  $\Delta t = 0.1 \tau$. $\xi$ and $\tau$ define the coherence length/time scales.}
\label{fig:time_series_example}
\end{figure}

In this paper, we investigate a driven-dissipative condensate in 1D subject to one-body loss $\Gamma_k \sim \lambda  k^2$.
We show that when perturbed from uniformity, this system exhibits a striking instability, best demonstrated by the example in Fig.~\ref{fig:time_series_example}.
Shown is the density profile of a condensate as a function of time, obtained by numerical simulation of the Gross-Pitaevskii equation (details to be explained below).
The condensate initially has a slight localized excess of particles.
The excess density begins to spread throughout the system, and the condensate appears to relax to a uniform state.
However, after a significant delay, the density quickly drops to zero in certain regions, forming fronts which move ballistically and eventually consume the entire condensate.

We show that the onset of instability can be attributed to the long-wavelength equation for the phase of the nearly-uniform condensate.
Whereas driven-dissipative condensates with $k$-independent loss are typically described by the Kardar-Parisi-Zhang (KPZ) equation~\cite{Kuramoto1984,Grinstein1993Temporally,Grinstein1996Conjectures,Altman2015TwoDimensional,Maghrebi_2017}, a well-known nonlinear diffusion equation, here we find an analogous nonlinear \textit{wave} equation which we refer to as ``dispersive KPZ''.
Little is known about the dispersive KPZ equation, at least in the physics literature, but a surprising feature of the latter is that generic solutions diverge in finite time~\cite{Escudero2007BlowUp}.
We show that this singularity corresponds to the sudden depletion of the condensate.

The dynamics following formation of the depleted regions can no longer be described by dispersive KPZ, for which solutions simply do not exist beyond the singularity time.
We thus derive a more general pair of hydrodynamic equations, and identify soliton solutions which accurately describe the shape and motion of the fronts seen in Fig.~\ref{fig:time_series_example}.
As will be clear, these solitons are exclusive to dissipative condensates, and in fact, their core size diverges in the limit of vanishing dissipation, $\lambda \rightarrow 0$.

\textit{Dissipative Gross-Pitaevskii equation.---}We consider a one-dimensional gas of particles with contact interactions and single-body loss $\Gamma_k\sim \lambda k^2$. 
Formally, the system is described by the quantum master equation ($\hbar = 1$)
\begin{align} \label{eq:Rydberg_polariton_master_equation}
\partial_t \rho &= -i \big( \hat H_{\textrm{eff}} \rho - \rho \hat H_{\textrm{eff}}^{\dag} \big) + \int \textrm{d}x \, \frac{\lambda}{m} (\partial_x \hat{\psi}) \rho (\partial_x \hat{\psi}^{\dag}), \\
 \label{eq:Rydberg_polariton_Hamiltonian}
&\hat H_{\textrm{eff}} = \int \textrm{d}x \left[ \frac{1 - i \lambda}{2m} (\partial_x \hat{\psi}^{\dag}) (\partial_x \hat{\psi}) + U \hat{\psi}^{\dag 2} \hat{\psi}^2 \right] ,
\end{align}
where $\rho$ is the density matrix of the system and $\hat{\psi}^{\dag}(x)$ creates a bosonic particle at position $x$.
Here $m$ is the mass and $U > 0$ governs the strength of interactions.

Following the standard procedure, e.g., as in Refs.~\cite{Kamenev2011,Sieberer2016Keldysh}, we first derive the semiclassical equation of motion for the condensate wavefunction $\psi(x)$, valid at large densities $\rho_0 \gg mU$:
\begin{equation} \label{eq:semiclassical_equation}
i \partial_t \psi + \frac{1 - i \lambda}{2m} \partial_x^2 \psi - 2U \big| \psi \big| ^2 \psi = 0.
\end{equation}
Equation~\eqref{eq:semiclassical_equation} is quite similar to the standard Gross-Pitaevskii (GP) equation, with the only difference being that the coefficient of the kinetic term is complex. Therefore, any spatial variation of the wavefunction leads to dissipation. We shall focus on the dynamics of a nearly-uniform condensate.
For concreteness, we use initial conditions of the form
\begin{equation} \label{eq:generic_initial_conditions}
\psi(x, 0) = \sqrt{\rho_0} \left( 1 + h e^{-\frac{x^2}{w^2}} \right) ^{\frac{1}{2}}.
\end{equation}
We have confirmed that the conclusions of this paper hold for other initial conditions as well (sinusoidal perturbations, random density/phase fluctuations, etc.).

The natural length scale of Eq.~\eqref{eq:semiclassical_equation} is the healing length $\xi \equiv \sqrt{1/mU\rho_0}$, and the natural time scale is $\tau \equiv m \xi^2$. 
The remaining dimensionless parameters are the dissipation strength $ \lambda$, the magnitude of the density perturbation $h$, and the width of the density perturbation $w/\xi$.

\begin{figure}[t]
\centering
\includegraphics[width=1.0\columnwidth]{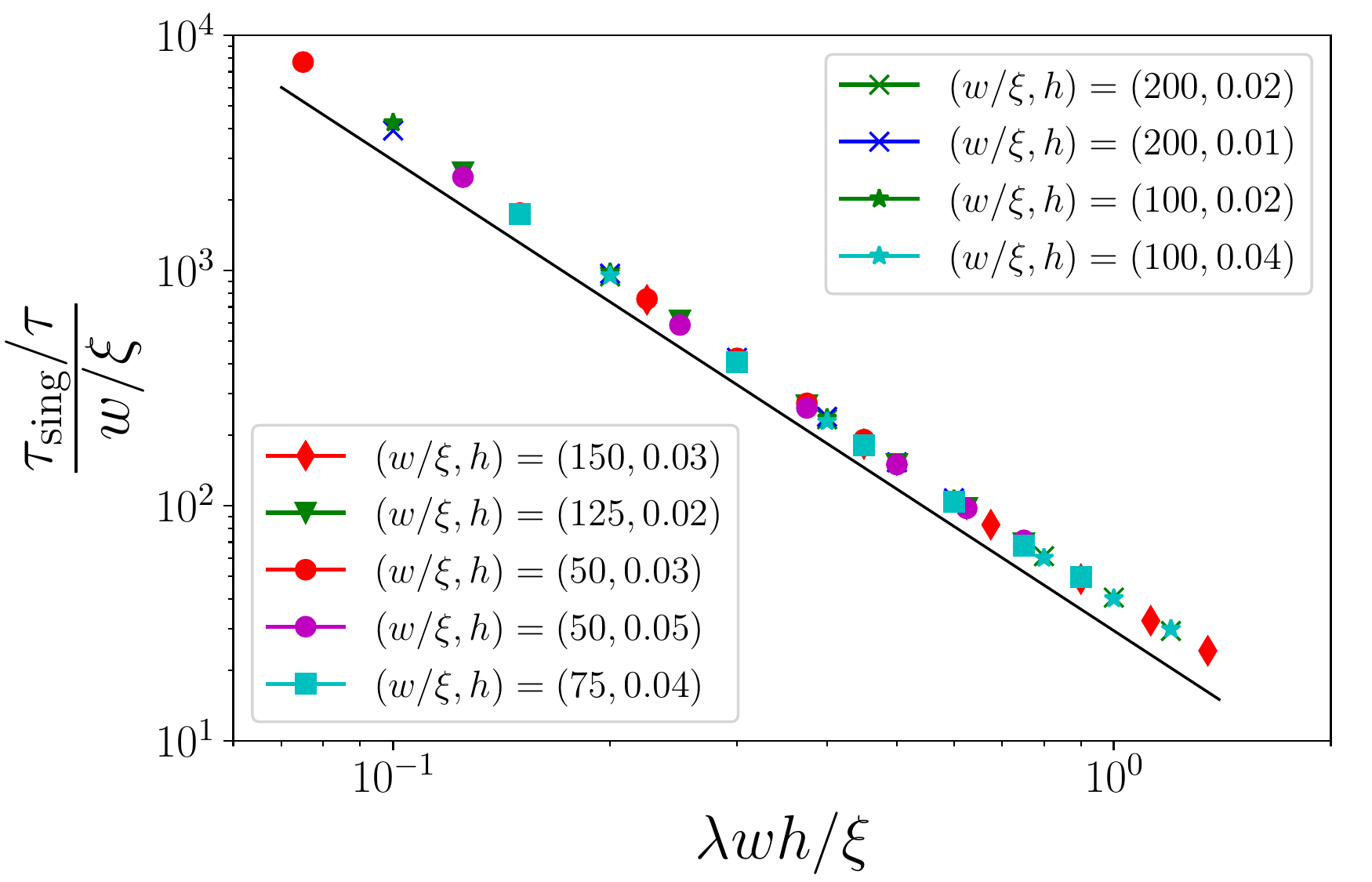}
\caption{Demonstration of the scaling form for the singularity time $\tau_{\textrm{sing}}$. The solid line is a power-law $z^{-2}$, drawn for comparison. $\tau_{\textrm{sing}}$ is plotted for Gaussian density perturbations of various heights $h$ and widths $w$ (see the inset), and various dissipation strengths $\lambda \in \{ 0.05, 0.1, 0.15, 0.2, 0.25, 0.3 \}$. System size is $L = 10000 \xi$.}
\label{fig:extinction_scaling}
\end{figure}

Fig.~\ref{fig:time_series_example}, showcased earlier, displays a representative simulation of Eq.~\eqref{eq:semiclassical_equation} using the initial profile in Eq.~\eqref{eq:generic_initial_conditions}.
The behavior is highly non-trivial---a prolonged period during which the condensate is nearly uniform is followed by the sudden appearance and subsequent spread of fully depleted regions.
We refer to the sudden depletion as a ``singularity''.
While the density profile is strictly analytic as a function of time, the long-wavelength equation derived below exhibits a genuine singularity which acts as a precursor to the condensate depletion. 

For concreteness, let us define $\tau_{\textrm{sing}}$ as the time when $\rho(x, t) \equiv |\psi(x, t)|^2$ first drops below $\rho_0/2$ at some position $x$, i.e., the first time at which $\textrm{min}_x \rho(x, t) < \rho_0/2$.
Figure~\ref{fig:extinction_scaling} plots $\tau_{\textrm{sing}}$ for multiple choices of $\lambda$ and initial conditions.
A clear scaling form is seen:
\begin{equation} \label{eq:extinction_time_scaling}
\frac{\tau_{\textrm{sing}}(\lambda, w, h)}{\tau} \sim \frac{w}{\xi} {\mathcal F} \left( \frac{\lambda w h}{\xi} \right),
\end{equation}
where the scaling function appears to fall off as ${\mathcal F}(z)\sim z^{-2}$ for $z\lesssim 1$.
Such algebraic dependence implies that the underlying instability is fundamentally different from nucleation, where a metastable state tunnels into a true equilibrium state, for which the decay rate would be exponentially suppressed at small fluctuations/perturbations.
The instability reported here is governed by a different mechanism that follows from the long-wavelength description of the condensate.

\textit{Dispersive KPZ equation.---}To derive the long-wavelength effective equation for the nearly-uniform condensate, starting from Eq.~\eqref{eq:semiclassical_equation}, we: i) write $\psi(x, t) = \sqrt{\rho_0 + \Delta \rho(x, t)} e^{i \theta(x, t)}$, assuming $\Delta \rho \ll \rho_0$; and ii) retain only the terms in the GP equation which are both lowest-order in $\Delta \rho / \rho_0$ and most relevant at long wavelengths.
The calculation is given in the Supplemental Material (SM)~\cite{suppRef}.
The end result is
\begin{equation} \label{eq:inertial_KPZ_equation}
\frac{1}{c^2}\partial_t^2 \theta =  \partial_x^2 \theta + \lambda  ( \partial_x \theta ) ^2,
\end{equation}
with $c \equiv \sqrt{2} \xi / \tau$ being a velocity scale that characterizes the ``speed of sound''.
The density variation in turn comes out to be $\Delta \rho = -({\rho_0 \tau}/{2}) \partial_t \theta$.

\begin{figure}[t]
\centering
\includegraphics[width=1.0\columnwidth]{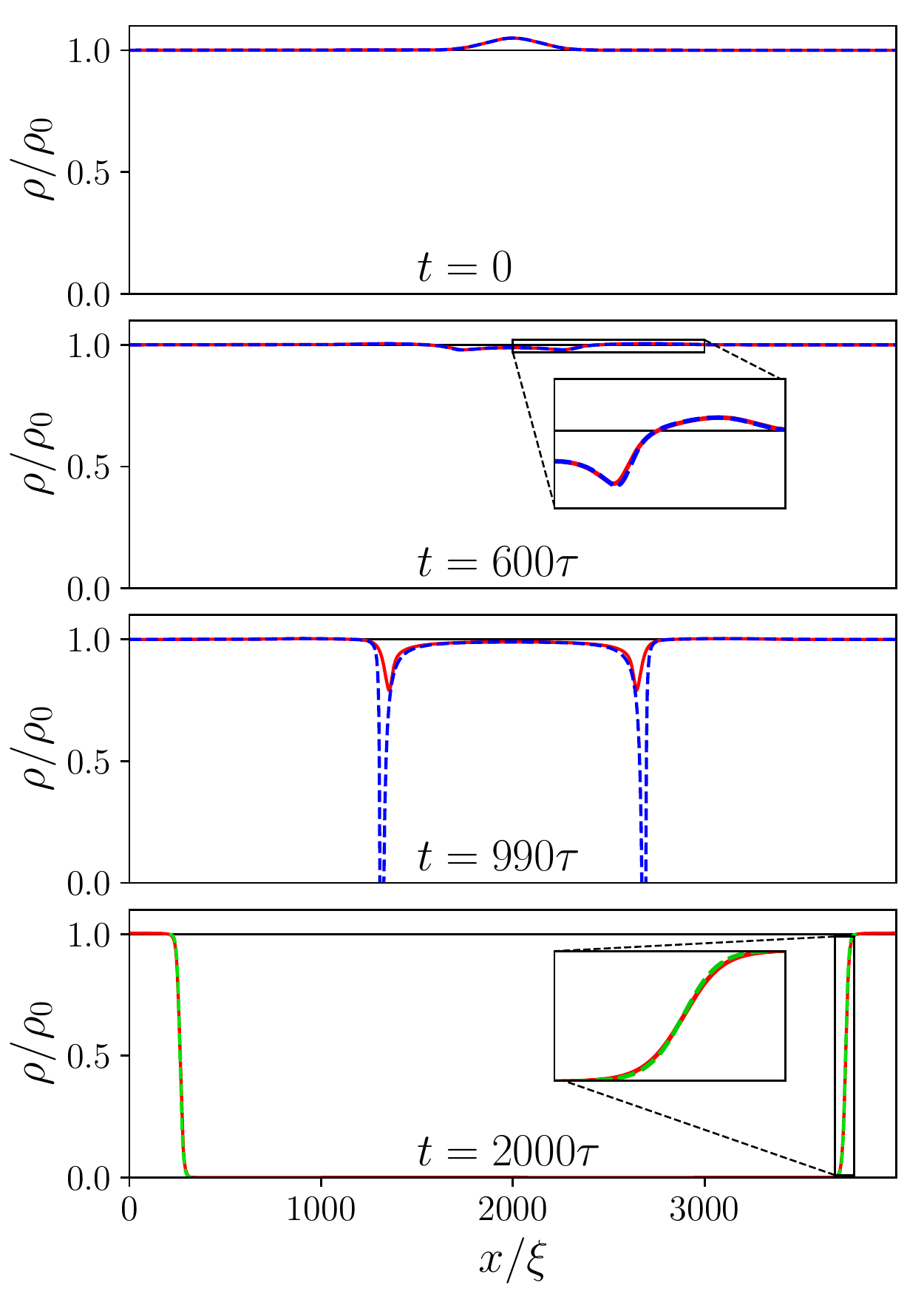}
\caption{Comparison of solution to GP equation (red lines) against solutions to approximate equations for time evolution of a Gaussian density perturbation. Times are indicated in each panel. (Upper three) Comparison to dispersive KPZ, shown in blue. (Bottom) Comparison to soliton given by Eqs.~\eqref{eq:hydro_soliton_solution} and~\eqref{eq:hydro_soliton_function}, shown in green, with the constant $z_0$ chosen to match the center of the front. Simulation details: dissipation strength is $\lambda = 0.4$, height of the initial Gaussian is $h = 0.05$, width is $w = 200 \xi$, spatial step size is $\Delta x = 0.2 \xi$, temporal step size is $\Delta t = 0.1 \tau$.}
\label{fig:linear_comparison}
\end{figure}

Equation~\eqref{eq:inertial_KPZ_equation} is quite similar to the (noiseless) KPZ equation which has emerged in generic dissipative condensates~\cite{Kuramoto1984,Grinstein1993Temporally,Grinstein1996Conjectures,Altman2015TwoDimensional}, except for the second time derivative on the left-hand side, which results in a wave-like equation with $c$ defining a causal ``light cone'' \cite{Escudero2007BlowUp}. 
Being a nonlinear wave equation, we refer to Eq.~\eqref{eq:inertial_KPZ_equation} as the ``dispersive KPZ'' equation.
Much less is known about dispersive KPZ than its diffusive counterpart~\cite{souplet1995nonexistence,Makarenko1997On,Liu2001Longtime,Orive2006Long}, but one established result is that under certain conditions, solutions to the dispersive KPZ equation---as well as a larger class of nonlinear hyperbolic equations---diverge in finite time~\cite{Escudero2007BlowUp}.
On physical grounds, this is due to the absence of any damping term such as $\partial_t \theta$ which could counteract the nonlinear growth.
We have confirmed this divergence through numerical simulation of Eq.~\eqref{eq:inertial_KPZ_equation}.

Figure~\ref{fig:linear_comparison} compares the solution of dispersive KPZ to the solution of the GP equation for a representative example.
We see that: i) the two agree extremely well for as long as $\Delta \rho / \rho_0$ is everywhere small; and ii) development of the singularity in dispersive KPZ coincides with the depletion of the condensate.
For this reason, we equate the singularity time with $\tau_{\textrm{sing}}$~\footnote{Unlike the singularity time in dispersive KPZ, which is a well-defined instant, our definition of $\tau_{\textrm{sing}}$ in the GP equation is somewhat arbitrary due to setting a threshold at $\rho_0/2$ (see the discussion above Eq.~\eqref{eq:extinction_time_scaling}). But in the limit where $\tau_{\textrm{sing}} \gg \tau$, using any other fraction instead of $1/2$ would change $\tau_{\textrm{sing}}$ only by a subleading amount.}.

The scaling form of $\tau_{\textrm{sing}}$ given in Eq.~\eqref{eq:extinction_time_scaling} then follows from the scaling of solutions to dispersive KPZ.
Suppose that, just as in the simulations above, initially $\theta(x, 0) = 0$ and $\Delta \rho(x, 0)= -\frac{\rho_0 \tau}{2} \partial_t \theta(x, 0)$ is of the form
\begin{equation} \label{eq:KPZ_initial_scaling}
\Delta \rho(x, 0) = \frac{\rho_0 h}{2} {\mathcal G} \left( \frac{x}{w} \right) ,
\end{equation}
for some dimensionless function ${\mathcal G}(y)$.
Defining $y \equiv x/w$, $s \equiv t c / w$, $\phi(y, s) \equiv \lambda \theta(x, t)$, the dispersive KPZ equation together with the initial conditions can be written as
\begin{equation} \label{eq:KPZ_scaled_equation}
\begin{gathered}
\partial_s^2 \phi = \partial_y^2 \phi + ( \partial_y \phi ) ^2, \\
\phi(y, 0) = 0, \quad \partial_s \phi(y, 0) = -\frac{\lambda w h}{\sqrt{2} \xi} {\mathcal G}(y).
\end{gathered}
\end{equation}
The only dimensionless parameter here is $\lambda w h / \xi$, hence the scaling form in Eq.~\eqref{eq:extinction_time_scaling}.
This gives further evidence for the applicability of dispersive KPZ~\footnote{For more complicated initial conditions, $\tau_{\textrm{sing}}$ will not obey so simple a scaling form.}.

Unfortunately, the dispersive KPZ equation does not have a general analytic solution (although a solvable special case is given in the SM~\cite{suppRef}).
Thus let us briefly discuss an analogous but simpler equation that exhibits similar features:
\begin{equation} \label{eq:first_order_nonlinear_wave_equation}
\partial_t \tilde \theta = \partial_x \tilde \theta + \lambda \tilde \theta^2,
\end{equation}
which, in dimensionless coordinates, describes a left-moving wave with an additional nonlinear term  (with $\tilde\theta$ roughly mimicking $\partial_x\theta$~\footnote{To see the analogy with dispersive KPZ, define (again setting $\tau = \xi = 1$ for simplicity) $\theta_1 \equiv \partial_t \theta, \theta_2 \equiv \partial_x \theta$. Dispersive KPZ can be written as the pair of first-order equations $\partial_t \theta_1 = \partial_x \theta_2 + \lambda \theta_2^2$, $\partial_t \theta_2 = \partial_x \theta_1$.}).
It is trivial to solve this equation by transforming to the frame moving alongside the wave: along the path $x(t) = x_0 - t$, Eq.~\eqref{eq:first_order_nonlinear_wave_equation} simply becomes $\textrm{d}\tilde \theta / \textrm{d}t = \lambda \tilde\theta^2$.
Thus the general solution is
\begin{equation} \label{eq:first_order_general_solution}
\tilde\theta(x_0 - t, t) = \frac{\tilde\theta_0}{1 - \lambda \tilde\theta_0 t},
\end{equation}
where $\tilde\theta_0 \equiv \tilde\theta(x_0, 0)$.
We see that, unless $\tilde\theta(x_0, 0)$ is everywhere negative, $\tilde\theta(x, t)$ will diverge in finite time, regardless of the precise shape of the initial condition.
The same phenomenon occurs in the setting of the dispersive KPZ equation.
Note that this behavior is much more drastic than a linear instability, where the amplitude would grow exponentially but nonetheless be finite at any finite time.

\textit{Hydrodynamic equations.---}For times greater than $\tau_{\textrm{sing}}$, the dispersive KPZ equation clearly cannot describe the evolution of the condensate.
Thus we derive a pair of hydrodynamic equations which no longer assume $\Delta \rho \ll \rho_0$, only requiring that the relevant length and time scales still be larger than $\xi$ and $\tau$, respectively.
We follow the standard procedure for quantum fluids by describing the wavefunction in terms of the density $\rho(x, t)$ and velocity field $v(x, t) \equiv \partial_x \theta(x, t) / m$~\cite{Pethick2002,Cazalilla2011One}.
The resulting hydrodynamic equations become (see the SM for details~\cite{suppRef})
\begin{align}
    \label{eq:hydro_continuity_equation}
&\partial_t \rho + \partial_x \big( \rho v \big) = -\lambda m \rho v^2, \\
\label{eq:hydro_Euler_equation}
&\partial_t v + v \partial_x v = -2mU^2 \partial_x \rho.
\end{align}
Equation \eqref{eq:hydro_continuity_equation} is the analogue of the continuity equation, with the additional feature that the density is depleted in regions of nonzero velocity.
Equation \eqref{eq:hydro_Euler_equation} is the standard Euler equation for an incompressible fluid, with the pressure given by $P(\rho) =  mU^2 \rho^2$ [hence the right-hand side can be written as $-\rho^{-1} \partial_x P(\rho)$]~\cite{Cazalilla2011One}. 

One can confirm by direct substitution that the above equations admit soliton solutions---$\rho(x, t) = \rho(x - u t)$, $v(x, t) = v(x - u t)$---for any velocity $u$ such that $|u| \geq c$ ($c$ being defined as before).
Supersonic solitons are likely unstable, therefore we focus on the case $u = c$, where the soliton moves rightward at the speed of sound.
In terms of $z \equiv x - c t$, we obtain~\cite{suppRef}
\begin{equation} \label{eq:hydro_soliton_solution}
\frac{\rho(z)}{\rho_0} = 1 + \frac{v(z)}{c} - \frac{v(z)^2}{2 c^2}, \; \;\;\; \frac{v(z)}{c} = f^{-1} \big[ \frac{\sqrt{2} \lambda}{3 \xi} (z_0 - z) \big] ,
\end{equation}
where $z_0$ is a constant which fixes the center of the soliton and $f^{-1}$ is the inverse of the function
\begin{equation} \label{eq:hydro_soliton_function}
\begin{aligned}
f(y) = \log{(-y)} & - \frac{2\sqrt{3} + 3}{6} \log{(\sqrt{3} - 1 + y)} \\
& + \frac{2\sqrt{3} - 3}{6} \log{(\sqrt{3} + 1 - y)}.
\end{aligned}
\end{equation}
Note that the density $\rho$ approaches $\rho_0$ as $z\to \infty$, while it vanishes as $z\to -\infty$. 
The fronts observed in our simulations of the GP equation agree well with Eq.~\eqref{eq:hydro_soliton_solution} (the left-moving fronts are easily related to the above by symmetry).
A representative comparison is shown in the bottom panel of Fig.~\ref{fig:linear_comparison}.

These solitons are quite different from those in the dissipation-free GP equation [$\rho\sim\rho_0 \tanh^2{(z/\sqrt{2}\xi)}$] \cite{Pethick2002}.
Most importantly, the dissipative solitons have a core size $\xi/\lambda$ (as opposed to simply $\xi$), which diverges in the limit of vanishing dissipation, $\lambda \rightarrow 0$.
This is consistent with the fact that these solitons originate from an instability which occurs only in the presence of quantum diffusion.

\textit{Physical realizations.---}Let us briefly comment on potential physical realizations of this phenomenon.
As noted above, one possible platform is Rydberg polaritons via electromagnetically-induced transparency (EIT), formed when an incoming photon hybridizes with a long-lived Rydberg state through a lossy intermediate state~\cite{Fleischhauer2000DarkState,Lukin2001Dipole,Friedler2005LongRange}.
At precisely zero momentum, the polariton is a superposition of Rydberg state and photon with exactly zero amplitude on the lossy state, and hence is essentially lossless.
The deviation from resonance at small but finite $k$ leads to the $k^2$ loss and results in a diffusion-like term~\cite{Peyronel2012Quantum}. 
Furthermore, at low energies, we can neglect scattering into other modes, leaving Eq.~\eqref{eq:Rydberg_polariton_Hamiltonian} as the effective many-body Hamiltonian.

While the interaction between polaritons is generically complex-valued as well, we have confirmed that it is possible to tune microscopic parameters so that the effective two-body loss rate vanishes while the one-body ($k^2$) loss remains significant; see the SM~\cite{suppRef}. 
Thus the instability reported here may be observable in Rydberg polariton systems, although the parameter regime in which $\tau_{\textrm{sing}} \gg \tau$ (where the ``singularity'' is sharpest) would necessitate a long atomic medium.
Running-wave cavities may provide a feasible alternative to the long free-space lengths.

An alternate realization could come from a 1D cloud of bosonic atoms driven by two coherent lasers under EIT condition.
With one beam orthogonal to the atomic gas and the other parallel, detuning (proportional to the atomic wave vector $k$) due to the Doppler shift leads to diffusion-like dynamics~\cite{Fleischhauer2005Electromagnetically}.
In order to ensure that the contact interaction does not itself cause losses, one would have to properly choose the states involved and tune interactions, e.g., with a magnetic field~\cite{Bloch2008Many}.
Finally, microcavity arrays \cite{Diehl2008Quantum,Houck2012OnChip} provide another platform where $k^2$ loss can be realized~\cite{Marcos_2012,Marino2016Driven}.
However, it may be challenging to engineer coherent interactions and diffusive terms simultaneously.

\textit{Conclusion.---}We have shown that 1D driven-dissipative condensates for which quantum diffusion is the dominant source of dissipation suffer from a peculiar instability to local density perturbations.
The condensate relaxes towards uniform density until a time $\tau_{\textrm{sing}}$---much larger than the natural timescale $\tau$---after which certain regions quickly deplete and form fronts which then spread throughout the condensate.
We have traced this behavior to the long-wavelength effective equation for the phase of the condensate, a nonlinear wave equation which we refer to as the ``dispersive KPZ'' equation.
Solutions to dispersive KPZ can diverge at finite times, and we have observed that the singularity in the long-wavelength description coincides with depletion of the condensate.
We have further derived a pair of hydrodynamic equations for the condensate that accurately describe the dynamics even beyond the onset of instability.
Interestingly, the fronts are described by non-standard soliton solutions that emerge solely due to dissipation. 

From a mathematical perspective, it has long been known that the solutions to nonlinear wave equations can diverge, or more generally become nonanalytic~\cite{John1990,Alinhac1995}.
It is interesting to note that whereas the divergence is often seen as an unphysical mathematical pathology, here it corresponds to a genuine physical phenomenon.
Coincidentally, Ref.~\cite{Escudero2007BlowUp} even comments: ``there is, to our knowledge, no direct application of [dispersive KPZ] to a physical problem''.
The situation discussed here---condensates undergoing quantum diffusion---provides such an application, the first such to our knowledge.

Many directions for future work remain.
First of all, it is desirable to go beyond the semiclassical limit and investigate the strongly-interacting quantum regime.
A step in this direction would be to include noise terms in Eqs.~\eqref{eq:semiclassical_equation} and~\eqref{eq:inertial_KPZ_equation}~\cite{Kamenev2011,Sieberer2016Keldysh}.
Most studies of the traditional KPZ equation do include a noise term, as it is the competition between noise and nonlinearity which leads to novel scaling properties~\cite{Kardar1986Dynamic,HalpinHealy1995Kinetic}, and so it is natural to ask whether dispersive KPZ has its own distinct scaling behavior.
Furthermore, the traditional solitons and hydrodynamic behavior of condensates have been well-studied~\cite{Pethick2002,Leggett2006,Cazalilla2011One}, whereas we have only scratched the surface of the present equations. 
Finally, further scrutiny of different physical realizations  is worthwhile.
While we have intentionally kept our analysis theoretical and abstract,  more systematic investigations are needed to assess the feasibility of any specific implementation.

\textit{Acknowledgments.}---The authors would like to thank J. V. Porto for informative discussions.
This research was performed while C.L.B.~held an NRC Research Associateship award at the National Institute of Standards and Technology.
P.B.~and A.V.G.~acknowledge funding by the AFOSR, AFOSR MURI, DoE ASCR Quantum Testbed Pathfinder program (award No. DE-SC0019040), U.S.~Department of Energy Award No. DE-SC0019449, DoE ASCR Accelerated Research in Quantum Computing program (award No. DE-SC0020312), NSF PFCQC program, and ARO MURI.  M.M. acknowledges support from NSF under Grant No. DMR-1912799,  the Air Force Office of Scientific Research (AFOSR) under award number FA9550-20-1-0073 as well as the start-up funding from Michigan State University.

\bibliography{main_biblio}

\newpage

\begin{widetext}

\section*{Supplemental Material to ``Singularities in nearly-uniform 1D condensates due to quantum diffusion''}

The contents of this Supplemental Material are as follows.
In Sec.~\ref{sec:dispersive_KPZ_derivation}, we show how the dispersive KPZ equation can be derived from the dissipative GP equation using two approximations: nearly-uniform density and long wavelengths.
In Sec.~\ref{sec:dispersive_KPZ_example}, we give a special case of a solution to dispersive KPZ which exhibits the finite-time singularity.
In Sec.~\ref{sec:hydrodynamic_derivation}, we derive the pair of hydrodynamic equations from dissipative GP by making only the long-wavelength approximation, and then identify soliton solutions to them.
Finally, in Sec.~\ref{sec:experimental_implementation}, we calculate the effective one-body and two-body loss rates in terms of microscopic parameters for Rydberg polariton systems, and demonstrate that it is possible to tune the two-body loss rate to zero.

\section{Derivation of dispersive KPZ from dissipative Gross-Pitaevskii} \label{sec:dispersive_KPZ_derivation}

Here we derive the dispersive KPZ equation, Eq.~(6) in the main text.
Starting from the dissipative Gross-Pitaevskii (GP) equation, reproduced here:
\begin{equation} \label{eq:dissipative_GP_equation}
i \partial_t \psi + \frac{1 - i \lambda}{2m} \partial_x^2 \psi - 2U \big| \psi \big| ^2 \psi = 0,
\end{equation}
let us first convert to a pair of equations for the density $\rho(x, t)$ and phase $\theta(x, t)$ via $\psi(x, t) \equiv \sqrt{\rho(x, t)} e^{i \theta(x, t)}$:
\begin{equation} \label{eq:GP_density_equation}
\partial_t \rho + \frac{1}{m} \partial_x \big( \rho \partial_x \theta \big) = \frac{\lambda}{m} \big( \sqrt{\rho} \partial_x^2 \sqrt{\rho} - \rho (\partial_x \theta)^2 \big) ,
\end{equation}
\begin{equation} \label{eq:GP_phase_equation}
\partial_t \theta + \frac{1}{2m} \big( \partial_x \theta \big) ^2 - \frac{1}{2m \sqrt{\rho}} \partial_x^2 \sqrt{\rho} + 2U \rho = \frac{\lambda}{2m \rho} \partial_x \big( \rho \partial_x \theta \big) .
\end{equation}
We next make two approximations:
\begin{itemize}
    \item The density can be written $\rho(x, t) = \rho_0 + \Delta \rho(x, t)$ with $\Delta \rho(x, t) \ll \rho_0$, and terms sub-leading in $\Delta \rho / \rho_0$ can be neglected.
    Note that merely by writing $\rho = \rho_0 + \Delta \rho$, Eq.~\eqref{eq:GP_phase_equation} acquires a constant term $2U \rho_0$ which we can trivially remove by taking $\theta \rightarrow \theta - 2U \rho_0 t$.
    \item All terms which are sub-leading at long wavelengths and low frequencies (in the sense to be defined momentarily) can be neglected.
\end{itemize}
The first approximation leads to the following (after defining $\xi \equiv \sqrt{1 / mU \rho_0}$, $\tau \equiv m \xi^2$ as in the main text):
\begin{equation} \label{eq:uniform_GP_density_equation}
\partial_t \Delta \rho + \frac{\rho_0 \xi^2}{\tau} \partial_x^2 \theta + \frac{\xi^2}{\tau} \big( \partial_x \Delta \rho \big) \big( \partial_x \theta \big) = \frac{\lambda \xi^2}{2 \tau} \partial_x^2 \Delta \rho - \frac{\lambda \rho_0 \xi^2}{\tau} \big( \partial_x \theta \big) ^2,
\end{equation}
\begin{equation} \label{eq:uniform_GP_phase_equation}
\partial_t \theta + \frac{\xi^2}{2\tau} \big( \partial_x \theta \big) ^2 - \frac{\xi^2}{4 \rho_0 \tau} \partial_x^2 \Delta \rho + \frac{2}{ \rho_0 \tau} \Delta \rho = \frac{\lambda \xi^2}{2 \tau} \partial_x^2 \theta + \frac{\lambda \xi^2}{2 \rho_0 \tau} \big( \partial_x \Delta \rho \big) \big( \partial_x \theta \big) .
\end{equation}
For the second, we heuristically consider $\partial_x$ to be of order $k \ll \xi^{-1}$, then neglect all terms which are necessarily sub-leading under \textit{only} the assumptions that $\xi k$ and $\Delta \rho / \rho_0$ are small (we are not making any assumptions for $\theta$).
For example, $\xi^2 \partial_x^2 \Delta \rho \sim \xi^2 k^2 \Delta \rho$ can be neglected compared to $\Delta \rho$ since $\xi k \ll 1$, and $\partial_x^2 \Delta \rho \sim k^2 \Delta \rho$ can be neglected compared to $\rho_0 (\partial_x \theta)^2 \sim \rho_0 k^2$ since $\Delta \rho \ll \rho_0$, but both $\Delta \rho / \rho_0$ and $\xi^2 \partial_x^2 \theta \sim \xi^2 k^2$ must be kept since we have not assumed anything about how $\Delta \rho / \rho_0$ compares to $\xi k$.
One then obtains
\begin{equation} \label{eq:long_wavelength_GP_density_equation}
\partial_t \Delta \rho = -\frac{\rho_0 \xi^2}{\tau} \partial_x^2 \theta - \frac{\lambda \rho_0 \xi^2}{\tau} \big( \partial_x \theta \big) ^2,
\end{equation}
\begin{equation} \label{eq:long_wavelength_GP_phase_equation}
\partial_t \theta = -\frac{\xi^2}{2\tau} \big( \partial_x \theta \big) ^2 + \frac{\lambda \xi^2}{2 \tau} \partial_x^2 \theta -\frac{2}{\rho_0 \tau} \Delta \rho.
\end{equation}
Now taking a time derivative of Eq.~\eqref{eq:long_wavelength_GP_phase_equation} together with Eq.~\eqref{eq:long_wavelength_GP_density_equation} yields a closed equation for $\theta$.
Also note that the right-hand side become small in the $k \rightarrow 0$, $\Delta \rho \rightarrow 0$ limit.
Thus we can treat $\tau \partial_t \sim \tau \omega$ as another small quantity, and neglect terms which are sub-leading by factors of $\tau \omega$.
We finally arrive at the equation for $\theta$:
\begin{equation} \label{eq:dispersive_KPZ_equation}
\partial_t^2 \theta = \frac{2 \xi^2}{\tau^2} \partial_x^2 \theta + \frac{2  \lambda \xi^2}{\tau^2} \big( \partial_x \theta \big) ^2.
\end{equation}
This is precisely the dispersive KPZ equation given in the main text [Eq.~(6)].
The excellent agreement in comparisons to the GP equation further confirms its validity.

\section{Example of a diverging solution to dispersive KPZ} \label{sec:dispersive_KPZ_example}

Here we give a specific solution to the dispersive KPZ equation which diverges in finite time.
Although the equation does not have a general analytic solution, it turns out that one family is simply parabolas with time-dependent coefficients:
\begin{equation} \label{eq:KPZ_parabola_ansatz}
\theta(x, t) = a(t) x^2 + b(t).
\end{equation}
The coefficients $a(t)$ and $b(t)$ are prescribed initial values and derivatives at $t = 0$, and the dispersive KPZ equation determines their subsequent values.

The fact that $\theta(x, t)$ diverges as $x \rightarrow \pm \infty$ makes Eq.~\eqref{eq:KPZ_parabola_ansatz} somewhat unphysical, but we will explain how these results imply singularities for more physical field profiles.
For now, merely note that $\theta(x, 0)$ is finite for any fixed value of $x$.
We shall show that $\theta(x, t) \rightarrow \infty$, still at fixed $x$, as $t$ approaches some finite $\tau_{\textrm{sing}}$.

Inserting Eq.~\eqref{eq:KPZ_parabola_ansatz} into Eq.~\eqref{eq:dispersive_KPZ_equation}, we find a closed pair of equations for $a(t)$ and $b(t)$:
\begin{equation} \label{eq:KPZ_parabola_equations}
\frac{\textrm{d}^2}{\textrm{d}t^2} a(t) = \frac{8 \lambda \xi^2}{\tau^2} a(t)^2, \quad \frac{\textrm{d}^2}{\textrm{d}t^2} b(t) = \frac{4 \xi^2}{\tau^2} a(t).
\end{equation}
The equation for $a(t)$, which gives the curvature of the $\theta$ field, is equivalent to the equation of motion for a particle in a cubic potential $U(a) \equiv -8 \lambda \xi^2 a^3 / 3 \tau^2$.
Thus unless the ``energy'' $E \equiv \frac{1}{2} (\textrm{d}a/\textrm{d}t)^2 + U(a)$ is exactly 0, the particle will ``roll downhill'' to $a = +\infty$.
The time required for $a$ to diverge is given by
\begin{equation} \label{eq:KPZ_parabola_divergence_time}
\tau_{\textrm{sing}} = \int_{a(0)}^{\infty} \textrm{d}a \frac{\textrm{sgn} \big[ \textrm{d}a / \textrm{d}t \big] }{\sqrt{2 \big( E - U(a) \big) }}.
\end{equation}
Since the integrand goes as $a^{-3/2}$ at large $a$, the integral converges and thus $\tau_{\textrm{sing}}$ is finite.

The singularity of the parabolic solution actually implies that a family of initially bounded $\theta$ profiles is singular as well.
The initial profile need only have an interval of sufficiently long length in which it is given by Eq.~\eqref{eq:KPZ_parabola_ansatz}.
This is because dispersive KPZ, like the linear wave equation, has a finite ``speed of light'': the value of $\theta(x_0, t_0)$ depends only on $\theta(x, t)$ for spacetime points in the (past) light cone $|x - x_0| \leq c(t_0 - t)$, where $c \equiv \sqrt{2} \xi / \tau$; a proof of such causal behavior can be found in Ref.~\cite{Escudero2007BlowUp}.
As a result, any $\theta(x)$ which is a parabola in a region of length $2L \equiv 2c \tau_{\textrm{sing}}$, where $\tau_{\textrm{sing}}$ is given by Eq.~\eqref{eq:KPZ_parabola_divergence_time}, will become singular in no more than time $\tau_{\textrm{sing}}$, regardless of the shape of $\theta(x, 0)$ outside the region.

\section{Derivation of hydrodynamic equations from dissipative Gross-Pitaevskii} \label{sec:hydrodynamic_derivation}

Here we first derive the hydrodynamic equations in the main text, Eqs.~(11) and~(12), and then construct soliton solutions to them.
For the derivation, we no longer assume $\Delta \rho \ll \rho_0$ but continue to make the long-wavelength approximation.
In particular, assume that spatial variations in $\rho$ and the velocity $v \equiv \partial_x \theta / m$ are on a significantly longer length scale than $\xi$.
Thus in Eq.~\eqref{eq:GP_density_equation}, written as
\begin{equation} \label{eq:GP_density_equation_rewrite}
\partial_t \rho + \partial_x \big( \rho v \big) = \frac{\lambda}{m} \sqrt{\rho} \partial_x^2 \sqrt{\rho} - m \lambda \rho v^2,
\end{equation}
the first term on the right-hand side, called the ``quantum pressure''~\cite{Cazalilla2011One}, can be dropped since it contains two spatial derivatives.
Equation~\eqref{eq:GP_phase_equation} in turn takes the form
\begin{equation} \label{eq:GP_phase_equation_rewrite}
\partial_t v + v \partial_x v - \frac{1}{2m^2} \partial_x \left( \frac{1}{\sqrt{\rho}} \partial_x^2 \sqrt{\rho} \right) + \frac{2U}{m} \partial_x \rho = \frac{\lambda}{2m} \partial_x \left( \frac{1}{\rho} \partial_x \big( \rho v \big) \right).
\end{equation}
Dropping terms containing two and three derivatives, we find Eqs.~(11) and~(12) in the main text.
Anticipating a solution where the density approaches a constant $\rho_0$ asymptotically and recalling $\xi \equiv \sqrt{1 / mU \rho_0}$, $\tau \equiv m \xi^2$, $c \equiv \sqrt{2} \xi / \tau$, these equations can be cast as
\begin{align} \label{eq:hydrodynamic_density_equation}
\partial_t \rho + \partial_x \big( \rho v \big) &= - \frac{\sqrt{2} \lambda}{\xi c} \rho v^2, \\
 \label{eq:hydrodynamic_velocity_equation}  
\partial_t v + v \partial_x v &= -\frac{c^2}{\rho_0} \partial_x \rho.
\end{align}

\begin{figure}[t]
\centering
\includegraphics[width=1.0\textwidth]{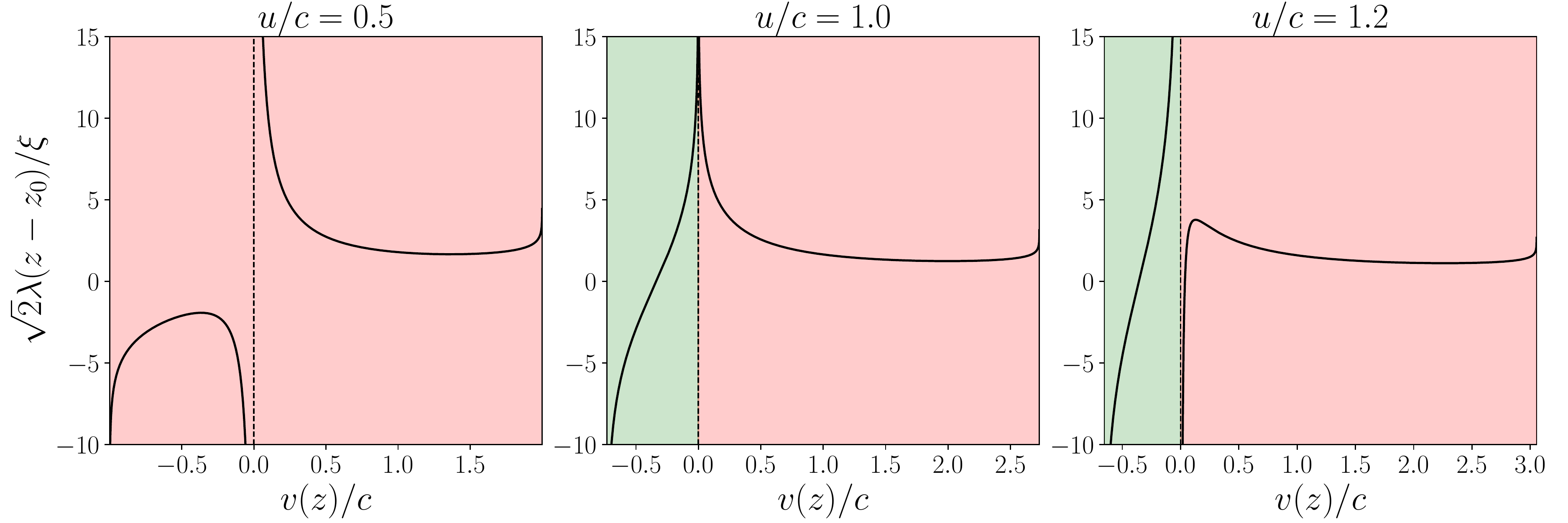}
\caption{Velocity profile of solitons, for three choices of $u/c$. Plotted is $z$ as a function of $v(z)$, in accordance with Eq.~\eqref{eq:hydrodynamic_velocity_soliton_solution}. The solution is valid only when $v(z)$ is single-valued for all $z$ (indicated by a green background). For certain $u$ and/or ranges of $v$, the solution is not single-valued and thus invalid (indicated by red).}
\label{fig:soliton_solutions}
\end{figure}

Soliton solutions to the above can be found by making the ansatz $\rho(x, t) = \rho(x - ut)$, $v(x, t) = v(x - ut)$, with $u$ to be determined.
Denote $z \equiv x - ut$.
Eq.~\eqref{eq:hydrodynamic_velocity_equation} becomes
\begin{equation} \label{eq:hydrodynamic_velocity_soliton_equation}
-u \frac{\textrm{d}v}{\textrm{d}z} + v \frac{\textrm{d}v}{\textrm{d}z} = -\frac{c^2}{\rho_0} \frac{\textrm{d}\rho}{\textrm{d}z},
\end{equation}
which can be integrated to give
\begin{equation} \label{eq:hydrodynamic_density_soliton_solution}
\rho(z) = \rho_0 \left( 1 + \frac{u}{c^2} v(z) - \frac{1}{2c^2} v(z)^2 \right) .
\end{equation}
To determine the integration constant, we have used the fact that in the region where the condensate is uniform and stationary, i.e., where $v(z) \rightarrow 0$, we should have $\rho(z) \rightarrow \rho_0$.
Since we are only interested in solutions with $\rho(z) \geq 0$, the allowed range of $v(z)$ is given by
\begin{equation} \label{eq:soliton_velocity_range}
u - \sqrt{u^2 + 2c^2} \leq v(z) \leq u + \sqrt{u^2 + 2c^2}.
\end{equation}

Substituting Eq.~\eqref{eq:hydrodynamic_density_soliton_solution} into Eq.~\eqref{eq:hydrodynamic_density_equation} gives
\begin{equation} \label{eq:hydrodynamic_density_soliton_equation}
\left( 1 - \frac{u^2}{c^2} + \frac{3u}{c^2} v(z) - \frac{3}{2c^2} v(z)^2 \right) \frac{\textrm{d}v}{\textrm{d}z} = -\frac{\sqrt{2} \lambda}{\xi c} \left( v(z)^2 + \frac{u}{c^2} v(z)^3 - \frac{1}{2c^2} v(z)^4 \right) .
\end{equation}
We have the partial fraction decomposition
\begin{equation} \label{eq:soliton_partial_fraction_decomposition}
\begin{aligned}
\frac{2c^2 - 2u^2 + 6uv - 3v^2}{2c^2 v^2 + 2u v^3 - v^4} =& \left( 2 + \frac{u^2}{c^2} \right) \frac{u}{c^2} \frac{1}{v} + \left( 1 - \frac{u^2}{c^2} \right) \frac{1}{v^2} \\
+& \, \frac{\sqrt{u^2 + 2c^2}}{\big( u + \sqrt{u^2 + 2c^2} \big) ^2} \frac{1}{v - u - \sqrt{u^2 + 2c^2}} - \frac{\sqrt{u^2 + 2c^2}}{\big( u - \sqrt{u^2 + 2c^2} \big) ^2} \frac{1}{v - u + \sqrt{u^2 + 2c^2}}.
\end{aligned}
\end{equation}
Thus Eq.~\eqref{eq:hydrodynamic_density_soliton_equation} is integrated to give
\begin{equation} \label{eq:hydrodynamic_velocity_soliton_solution}
\begin{aligned}
\frac{\sqrt{2} \lambda}{\xi c} \big( z - z_0 \big) =& \left( 1 - \frac{u^2}{c^2} \right) \frac{1}{v(z)} - \left( 2 + \frac{u^2}{c^2} \right) \frac{u}{c^2} \log{\big| v(z) \big|} \\
-& \, \frac{\sqrt{u^2 + 2c^2}}{\big( u + \sqrt{u^2 + 2c^2} \big) ^2} \log{\Big| v(z) - u - \sqrt{u^2 + 2c^2} \Big|} + \frac{\sqrt{u^2 + 2c^2}}{\big( u - \sqrt{u^2 + 2c^2} \big) ^2} \log{\Big| v(z) - u + \sqrt{u^2 + 2c^2} \Big|},
\end{aligned}
\end{equation}
where $z_0$ is the integration constant.
Inverting Eq.~\eqref{eq:hydrodynamic_velocity_soliton_solution} gives the velocity profile $v(z)$.

Not every choice of $u$ yields a valid solution, however.
In order for Eq.~\eqref{eq:hydrodynamic_velocity_soliton_solution} to be invertible and $v(z)$ to be defined for all $z$, the curve $z(v)$ must be monotonic and range from $-\infty$ to $\infty$ over an interval of $v$.
Figure~\ref{fig:soliton_solutions} gives some examples using $u > 0$ (the case $u < 0$ is related by symmetry---take $v \rightarrow -v$, $z \rightarrow -z$ to obtain the solution for $-u$).
We see that $u = c/2$ does not have any valid solution---we have confirmed that this is the case for any $|u| < c$. 
Furthermore, the cases $u = c$ and $u = 6c/5$ only have valid solutions for $v < 0$. 
One can confirm that the same is true for any $u > c$ (e.g., by setting $\textrm{d}z/\textrm{d}v = 0$ in Eq.~\eqref{eq:hydrodynamic_density_soliton_equation}---there will be two positive solutions for any $u > c$, hence the $z(v)$ curve is qualitatively similar to the rightmost panel of Fig.~\ref{fig:soliton_solutions}).

To summarize, soliton solutions exist only for $|u| \geq c$ and with the condensate moving in the opposite direction ($v(z) < 0$ for $u > 0$ and $v(z) > 0$ for $u < 0$).
In numerical simulations, however, solitons with $|u|>c$ do not appear, which can be argued on several grounds.
First, the long-wavelength dispersive KPZ equation is causal \cite{Escudero2007BlowUp}, hence the wavefront cannot cut into the condensate -- in a region still described by the latter equation -- at a speed faster than $c$. 
Second, such supersonic solitons, while emerging in many (even relativistic) field theories~\cite{wazwaz2002partial}, are not expected to emerge for physical initial conditions (for example, when the perturbation away from a uniform condensate is confined to a finite region).
Furthermore, even if they do emerge, they are likely to be unstable due to mechanisms such as the Landau criterion for the onset of dissipation in a superfluid~\cite{Pethick2002}.

\section{
Effective contact interactions between dark-state Rydberg polaritons
} \label{sec:experimental_implementation}

Here we calculate the effective one-body and two-body loss rates for a Rydberg polariton system given by the Hamiltonian
\begin{equation} \label{eq:Rydberg_polariton_Hamiltonian}
H = \int \textrm{d}z \begin{pmatrix} \psi_e(z) \\ \psi_p(z) \\ \psi_s(z) \end{pmatrix}^{\dag} \begin{pmatrix} -ic \partial_z & g & 0 \\ g & -\Delta & \Omega \\ 0 & \Omega & 0 \end{pmatrix} \begin{pmatrix} \psi_e(z) \\ \psi_p(z) \\ \psi_s(z) \end{pmatrix} + \frac{1}{2} \int \textrm{d}z \textrm{d}z' \frac{C_6}{\big| z - z' \big| ^6} \psi_s(z)^{\dag} \psi_s(z')^{\dag} \psi_s(z') \psi_s(z),
\end{equation}
where $\psi_e(z)$, $\psi_p(z)$, and $\psi_s(z)$ are the bosonic field operators for respectively a photon, atomic $p$ state, and atomic Rydberg state at position $z$.
$c$ is the speed of light, $g$ is the atom-photon coupling, $\Delta \equiv \delta + i \gamma$ is the complex detuning which takes into account the finite linewidth $2 \gamma$ of the $p$ state, and $2 \Omega$ is the control field Rabi frequency.
We shall show that it is possible to tune parameters so that the two-body loss rate vanishes while the one-body loss rate (going as $k^2$) remains finite.

We shall refer to Refs.~\cite{Gullans2016,Bienias2014Scattering}, and only present the relevant equations here.
First, for $g\gg |\Delta|,\Omega,\gamma$ and small $k$, we can focus on the middle branch (the dark state) resulting from diagonalization of the non-interacting part of the Hamiltonian.
The effective (complex) mass $m$ of the dark-state polariton is given by
\begin{equation}
m=-\frac{g^4}{2\Delta c^2\Omega^2},
\end{equation}
hence in the notation of the main text, the one-body ($k^2$) dissipation strength is simply $\lambda = \gamma / |\delta|$ for $\delta < 0$.
The effective interaction potential between polaritons is
\begin{equation}
U(r)=\frac{1}{\chi} \frac{1}{\frac{|\chi C_6|}{\chi C_6} r^6/r_b^6-1},
\end{equation} 
where the blockade radius is $r_b \equiv |C_6\chi|^{1/6}$, and 
\begin{equation}
\chi \equiv \frac{1}{2\Delta}-\frac{\Delta}{2\Omega^2}.
\end{equation}
    
For coherent interactions, the two-body problem can be solved analytically in the limit $\kappa r_b |\Delta/2\Omega^2\chi| \ll 1$~\cite{Bienias2014Scattering,Gullans2016}.
The analysis carries over to the more general dissipative case considered here.
Assuming weak interactions, we can approximate the effective interaction potential by a square well of width $r_b$ and depth $u_0^2$ chosen to match 
\beq
u_0^2 r_b=-\int_0^\infty \text{d}r\,  m U(r) 
=
-\frac{\pi}{3} r_b\kappa^2 \frac{ | \Delta | ^2 \left(\frac{\Delta }{\Delta ^2-\Omega ^2}\right)^{5/6}}{\Delta \Omega ^{1/3}  \sqrt[6]{\left| \frac{1}{\Delta }-\frac{\Delta }{\Omega ^2}\right| }},
\eeq
where $\kappa=g^2/c|\Delta|$.
In this case, we can find a complete analytic solution to the scattering phase for all $k$~\cite{Barlette2000}.
The contact interaction between polaritons in a dilute regime can be characterized by the strength $-1/ma$ (denoted $U$ in the main text) with 
\beq
a=r_b +\frac{1}{u_0 \tan(u_0r_b)}.
\eeq
More generally, however, one must solve the two-body problem numerically.
In either event, by choosing appropriate values for $\kappa r_b$ and $\delta$, one can hope to access a regime in which $ma\in \mathbb{R}$, i.e., interactions are lossless. 

\begin{figure}[t]
\centering
\includegraphics[width=0.5\columnwidth]{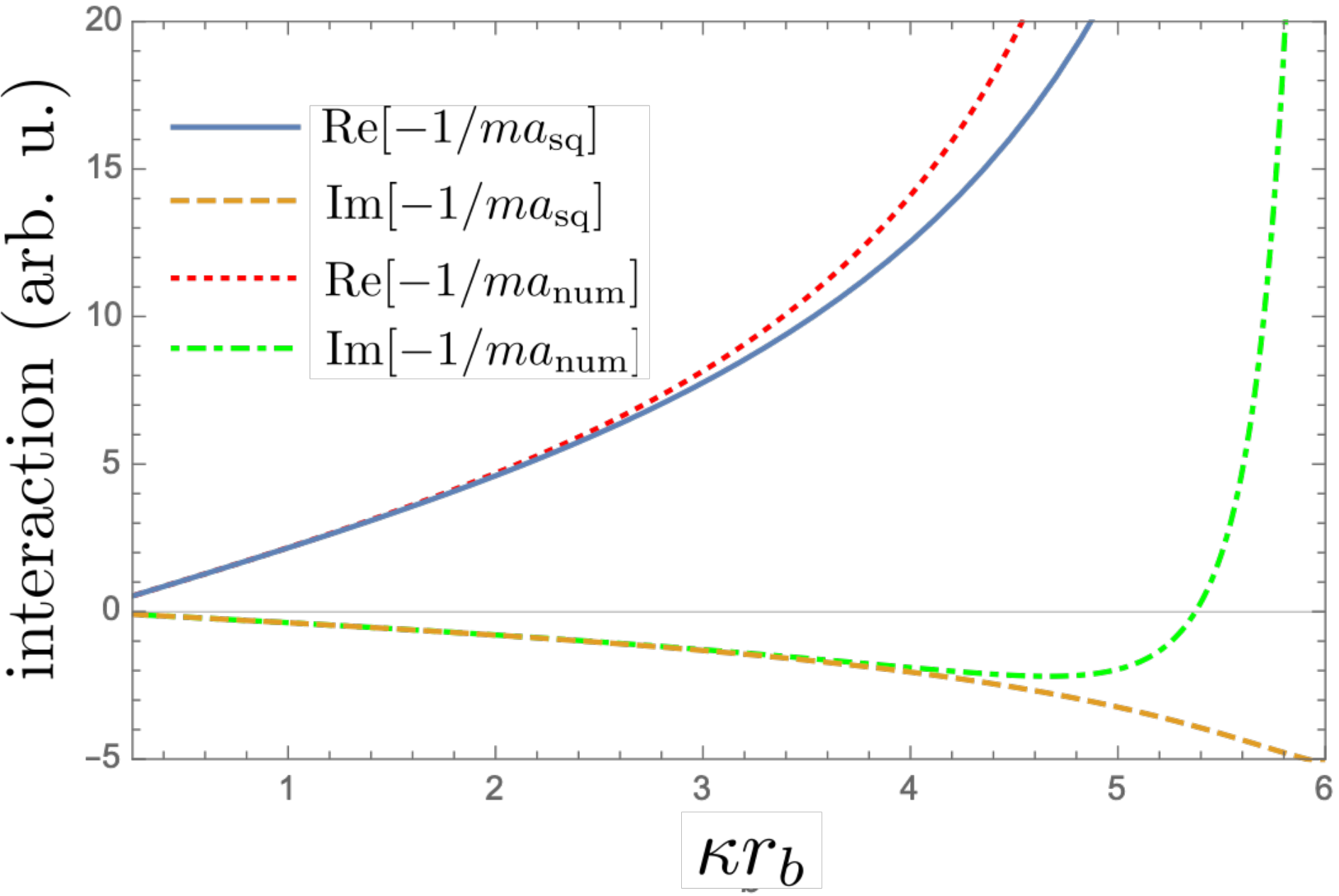}
\caption{Strength of contact interactions $-1/ma$ (denoted $U$ in the main text) between two dark-state polaritons for $\delta/\gamma = -10$, $\Omega/|\Delta|=6$, and $C_6>0$.
Solid blue and dashed orange curves are respectively the real and imaginary parts within the square-well approximation, while dotted red and dot-dashed green curves are those of the full numerical solution.
Using the numerical solution, we see a zero-crossing of Im$(1/ma)$ for $\kappa r_b\approx 5.5$.
}
\label{fig:scattering}
\end{figure}

We have found that such a regime exists, and that furthermore: i) this occurs when $\Omega > |\Delta|$, and thus the effective interactions are repulsive~\cite{Bienias2014Scattering}; and ii) $|\delta| \gg \gamma$, and thus the effective mass has a small (but nonzero) imaginary part.
These are the ideal conditions for the phenomena observed in the main text.

Figure~\ref{fig:scattering} shows a representative example, where $\delta/\gamma = -10$, $\Omega/|\Delta| = 6$, and $C_6 > 0$.
We see the desired zero-crossing of Im$(1/ma)$ for $\kappa r_b\approx 5.5$.
We have also compared the analytic square-well results ($-1/ma_{\textrm{sq}}$) against the numerical solution of the two-polariton scattering problem ($-1/ma_{\textrm{num}}$). 
As expected, the two agree at small $\kappa r_b$. The full numerical solution is required to identify the zero-crossing of the imaginary part.

\end{widetext}

\end{document}

%% file: commands2018.tex
   \setul{0.5ex}{0.1ex}
       \definecolor{Blue}{rgb}{0,0.0,1}
    \setulcolor{Blue}

\newcommand\reduline{\bgroup\markoverwith
{\textcolor{orange}{\rule[-.5ex]{2pt}{0.4pt}}}\ULon}

\newcommand{\remove}[1]{}

\graphicspath{ {./Figures/} }

\newcommand{\blst}{\begin{lstlisting}}
\newcommand{\est}{\end{lstlisting}}

\newcommand{\bit}{ \begin{itemize}{}  }
\newcommand{\eit}{\end{itemize}}

\newcommand{\be}{\begin{enumerate} \itemsep -4pt  }
\newcommand{\ee}{\end{enumerate}}

\newcommand{\bi}{\begin{itemize}   } 
\newcommand{\ei}{\end{itemize}}

\newcommand{\bma}{\begin{math}} 
\newcommand{\ema}{\end{math}}

\newcommand{\bc}{\begin{columns}} 
\newcommand{\ec}{\end{columns}}

\newcommand{\bbl}{\begin{block}} 
\newcommand{\ebl}{\end{block}}

\newcommand{\bflsh}{\begin{flashcard}}
\newcommand{\eflsh}{\end{flashcard}}

\newcommand{\bfl}[2]{\begin{flashcard}{#1} {#2} \eflsh}

\newcommand{\beq}{\begin{equation} \vspace{-0em}} 
\newcommand{\eeq}{\vspace{-0.em} \end{equation}}

\newcommand{\beqs}{\begin{equation*}}
\newcommand{\eeqs}{\end{equation*}}

\def\beqa{\begin{eqnarray}}
\def\eeqa{\end{eqnarray}}

\newcommand{\beal}{\begin{align}}
\newcommand{\eeal}{\end{align}}

\newcommand{\bvr}{\begin{verbatim}}
\newcommand{\evr}{\end{verbatim}}